\documentclass[a4paper]{mem}
\usepackage{natbib}
\usepackage{graphicx}
\usepackage[a4paper]{hyperref}
\idline{73}{23}
\def\gtsima{$\; \buildrel > \over \sim \;$}
\def\ltsima{$\; \buildrel < \over \sim \;$}
\def\prosima{$\; \buildrel \propto \over \sim \;$}
\def\gsim{\lower.5ex\hbox{\gtsima}}
\def\lsim{\lower.5ex\hbox{\ltsima}}
\def\simgt{\lower.5ex\hbox{\gtsima}}
\def\simlt{\lower.5ex\hbox{\ltsima}}
\def\simpr{\lower.5ex\hbox{\prosima}}

\def\ie{{\frenchspacing i.e. }}
\def\eg{{\frenchspacing e.g. }}

\def\HI{\hbox{H~$\scriptstyle\rm I\ $}}

\def\HeII{\hbox{He~$\scriptstyle\rm II\ $}}

\def\CIV{\hbox{C~$\scriptstyle\rm IV\ $}}
\def\SIV{\hbox{Si~$\scriptstyle\rm IV\ $}}

\def\nHI{{\rm HI}}

\def\nHeII{{\rm HeII}}

\def\vac#1{}
\begin{document}
   \title{Metagalactic Ultraviolet Flux from Cosmic Structure Formation
}

   \author{F. Miniati 
}


   \institute{
Max-Planck-Institut f\"ur Astrophysik,
Karl-Schwarzschild-Str. 1, 85740, Garching, Germany 
             }

   \abstract{
The contribution to the ultraviolet background (UVB) from thermal 
emission due to gas shock heated during cosmic structure formation
is assessed with an updated version of Press-Schechter (Sheth \& 
Tormen 1999) formalism.
The calculation is consistent with empirical
estimates based on the observed properties of galaxies and the observed
cosmic star formation history.  The bulk of the
radiation turns out to be produced by objects in the mass range $10^{11-13}
M_\odot$, \ie large galaxies and small groups. 
When compared to more conventional components (QSOs, stellar) 
it is found that near 1 Ry, thermal emission is comparable to 
stellar contributions and amounts to about 10 \%, 20 \% and 35
\% of the total flux at redshifts of 3, 4.5 and higher,
respectively; more importantly, near the ionization threshold for
\HeII, the thermal contribution is 
comparable to the QSO intensity already at redshift $\sim 3$ and
dominates at redshifts above 4. In addition, thermal photons alone are
enough to produce and sustain \HeII reionization already at $z\approx
6$. Finally, the observed Gunn-Peterson effect at high
redshifts ($3\leq z \leq6$) 
constrains the escape fraction of ionizing photons from
galaxies to less than a few percent.  
   }
   \authorrunning{F. Miniati et al.}
   \titlerunning{UV Flux from Structure formation}
   \maketitle
\section{Introduction}
Neutral hydrogen in the intergalactic medium (IGM) produces a forest
of resonant Ly$\alpha$ absorption lines in the spectra of
high-redshift quasars. The connection of these features 
to the structure formation process 
has been now firmly established using
N-body/hydrodynamic numerical simulations
\citep{CenApJL1994,ZhangApJ1995,MiraldaEscudeApJ1996}.  
There is consensus that the observed IGM temperature
results from a balance between photo-ionization heating and
adiabatic cooling due to the Hubble expansion. 
Such photo-heating is provided by the extragalactic
ultraviolet background (UVB) whose nature, origin and 
evolution have, therefore, been subject to considerable 
investigation.

At low redshifts, the ionization balance is consistent with a pure
power-law UVB. Traditionally QSOs have been considered as the
main sources of ionizing photon. However, a number of recent
results are at odds with this.
\citet{KimA&A2001} find the break in the
redshift evolution of absorbers at lower redshifts than
predicted by numerical simulations using a standard QSO
ionizing background \citep{hama96}, hinting at an incomplete
description of the UVB.  This led \citet[][B01
hereafter]{BianchiA&A2001} to recompute the UVB as a 
superposition of contributions from QSOs and galaxies. 

Similarly, high resolution simulations using a
cold dark matter model and a standard QSOs
ionizing background, produce Ly-$\alpha$
forest lines with a minimum width significantly below that observed
\citep{TheunsMNRAS1998a,BryanApJ1999}.
This has prompted several suggestions for 
additional heat sources, including
photo-electric dust heating \citep{NathMNRAS1999},
radiative transfer effects \citep{AbelApJ1999},
or Compton heating by X-ray background photons 
\citep{MadauApJL1999}. 

The situation around redshift $z\sim 3$ is more complicated.
There is a general trend for the optical depth at the ${\rm He}^+$ edge
to increase with redshift \citep[e.g.][]{Heap00}. \citet{Songaila1998} 
reports an abrupt 
change of the $\CIV/\SIV$ ratio at $z\approx 3$, often interpreted 
as a hardening of the ionization spectrum due to a sudden 
\HeII reionization. 
This is corroborated by the detection of patchy \HeII Ly$-\alpha$
absorption at similar redshifts.  
However, more recent VLT/UVES 
\citep{KimA&A2002} and Keck/HIRES \citep{bosara03}
studies find no such discontinuity
around $z=3$, and even suggest that $\CIV/\SIV$ is not a good
indicator of the \HeII ionization state.

Here we report on recent findings about
another source of UVB ionizing photons, namely
thermal emission from shock-heated gas in collapsed cosmic structures.
We show that the ionizing photons emitted by this
process make a non negligible fraction of the metagalactic flux, 
that the resulting spectrum is hard and
copious \HeII ionizing photons are produced. In fact this process may
well dominate the production of such hard photons at $z> 4$.

\section{Model} \label{model.se}
We consider the mean ultraviolet ionizing flux produced 
by QSO, stellar and thermal components.
After estimating the mean volume emissivity of each component as a
function of redshift, as described below, we solve for the 
cosmological radiative transfer equation \citep{peebles93}.
The {\it effective} optical depth, $\tau^\mathrm{eff}$, due to the 
presence of intervening neutral gas,
is computed by assuming a distribution of absorbers 
as a function of \HI column density
and redshift \citep{ParesceApJ1980}.
The dependence on $N_{\nHI}$
is determined from counts of Ly$\alpha$ absorption
lines in QSOs spectra; a power-law evolution with redshift
such that  $\tau^\mathrm{eff} \sim (1+z)^\gamma$, is further assumed 
\citep{ZuoA&A1993}.
We use $\gamma \simeq 3.4$ in 
the redshift range $1.5 < z < 4$ \citep{KimA&A2001},
and, based on the strong evolution for $\tau^\mathrm{eff}$ as
implied by recent spectra of high redshift QSOs
\citep{BeckerAJ2001,FanAJ2002} we adopt $\gamma=5.5$ for $z>4$.

The contributions of QSOs and stars to the ionizing UV 
background adopted here are similar to those in B01.
Basically the QSO emissivity assumes a luminosity function
that follows the double power-law model of \citet{BoyleMNRAS1988}. 
For $z\leq3$, we adopt the parameters given in 
\citet{BoyleMNRAS2000}. At $z>3$, we include the exponential 
decline suggested by \citet{FanAJ2001}.
For the QSO spectrum in the ionizing UV range we use
a simple power law, $j(\nu)\propto \nu^{-1.8}$ \citep{ZhengApJ1997}.

For the stellar component, we assume a star formation rate that 
is constant from high redshifts to $z\approx1$, then rapidly 
decreases to local values, as indicated 
by galaxy surveys in the rest-frame non-ionizing UV
\citep{MadauApJ1998,SteidelApJ1999}. Synthetic galactic spectra 
(\citealt{BruzualApJ1993}; version 2001) are then
used to calculate the emissivity of ionizing UV photons as a function 
of $z$. The absorption of radiation 
by the galaxy interstellar 
medium is modeled by a redshift-independent value for 
$f_\mathrm{esc}$, the fraction of Ly-continuum photons that 
can escape into the IGM.

Finally, for the emissivity from shocked intergalactic gas we assume
such gas to be thermalized in virialized dark matter halos, which we
describe with a Press-Schechter prescription \citep{shto99}. The dark
matter density in virialized objects is computed according to the
spherical collapse model; the ratio of baryons to dark matter is taken
as the universal value, $f_b\simeq 0.15$.  The spectral thermal
emissivity of the shocked gas in each halo is thus computed through
the code by \citet[][version 1992; collisional regime is a good
approximation for the plasma conditions]{rasm77}; we then 
estimate the thermal
volume emissivity by adding up the contributions from all collapsed
structures. We account for the finite cooling time of a halo, by
introducing a term ensuring that the radiated energy does not exceed
the thermal energy of the system.  Finally, we take into account
feedback effects by computing the increase in the cooling time caused
by the injection of additional, feedback energy.

More details are presented in \cite{mfwb04}. We note that cooling 
process associated with thermal emission investigated here
also leads to galaxy/star formation \citep{whre78}. In fact, 
\citet{mfwb04} also show that
the estimated flux presented here, 
based on a Press-Schechter model, is in remarkable
agreement with an empirical
estimate based on the observed cosmic star formation history
and the distribution of stellar mass as a function of halo 
virial temperature, as reconstructed from SDSS data from
\cite{kauffmannetal03}.
\section{Results}

\subsection{Photo-ionization Rates}
\label{photrat.se}
%

   \begin{figure*}
   \centering
   \resizebox{\textwidth}{!}
        {\includegraphics[clip=true]{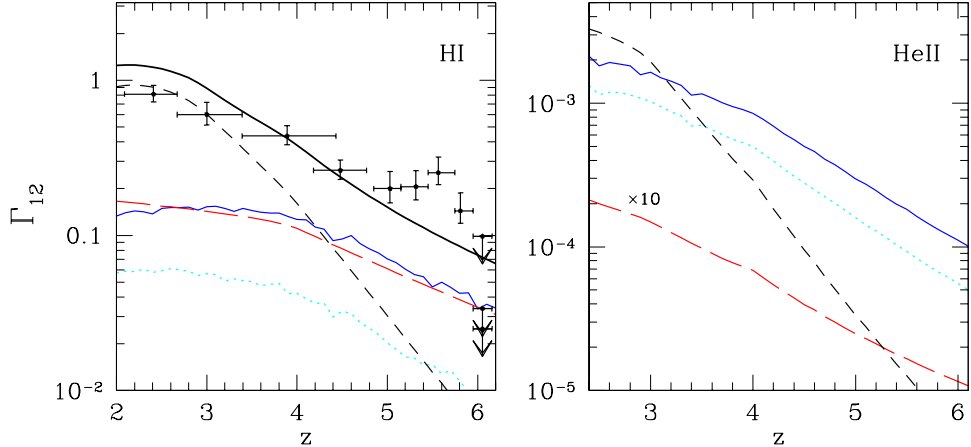}}
      \caption{Left: Photo-ionization rates defined in eq. 
\ref{phion.eq} as a function of redshift for ionization of 
\HI due to emission from: QSO (dash),
stellar for $f_{esc}=1\%$ (long dash) and 
shocked IGM with feedback effects (solid) and without (dot).
The thick solid line is the total considering the feedback case
(or, alternatively, the no-feedback case and $f_{esc}=2\%$). 
The data points are from 
\citet{McDonald2001ApJ} and \citet{FanAJ2002} after correction
for our cosmological model. 
Right: same as for left panel but now for coefficients 
relative to \HeII (right). The stellar component (long dash) has
now been multiplied by a factor 10 for visualization purposes.
              }
         \label{gammi.fig}
   \end{figure*}
%
%
Fig. \ref{gammi.fig} shows the evolution as a function of redshift,
of the 
photo-ionization rates in units of $10^{-12}$ s$^{-1}$ defined as
\begin{equation} \label{phion.eq}
\Gamma_{12} (z)=
\frac{4\pi}{10^{-12}\, {\rm s}^{-1}} \, \int_{\nu_s}^\infty \; 
\sigma_s(\nu) \; \frac{J(\nu,z)}{h_p\nu} \; d\nu
\end{equation}
for the various UV radiation components discussed above,
together with measured values 
inferred from the observed Gunn-Peterson effect in high-z QSO 
spectra \citep{McDonald2001ApJ,FanAJ2002}.

The left panel of Fig. \ref{gammi.fig} is relative to 
ionization of \HI and contains a number of important features. 
First, although the emission from QSOs 
is able to produce the ionizing flux observed at $z\sim 2-3$,
it falls short at higher redshifts, a well known fact. In our 
formulation, it results from the assumed rapid decline of the QSO 
number density for $z>3$, as derived from the SDSS 
\citep[\S \ref{model.se},~][]{FanAJ2001}.

Second, a comparison of the measured values of
$\Gamma_{12}$ and the stellar ionization rates assuming 
$f_{esc}=1\%$ (dash curve) implies, according to our model, that 
$f_{esc}$ is smaller than a few \%. 
 Estimates of the UV background from the proximity
effect are known to be larger than those obtained via theoretical models
of the IGM opacity (\ie the work of \citeauthor{FanAJ2002} we are using
here), probably because of a bias of the QSO distribution 
towards the denser environments \citep{SchirberApJ2002} 
or because of systematic errors
due to line blending \citep{ScottApJS2000}. 

Finally, the right panel of Fig. \ref{gammi.fig} shows 
the \HeII ionization rates. According to the plot, above
4 Ry stellar emission is thoroughly negligible (independent 
of $f_\mathrm{esc}$) whereas thermal emission 
is comparable to QSOs at $z\sim 3$ but completely dominates 
the radiation flux at higher redshifts. 
This result is very important in terms of the IGM evolution and 
has not previously been noticed.
It depends only weakly on feedback, but it does
assume an escape fraction of the thermal photons 
from collapsed halos of order 1. As 
discussed in \cite{mfwb04} these conditions should be ensured for 
collisionally ionized gas within halos of virial temperature above
$10^6$ K, that is for the halos that generate most of the thermal
emission.
%
%
%
%
%
\subsection{Softness parameter}
%
   \begin{figure}
        \includegraphics[width=7cm]{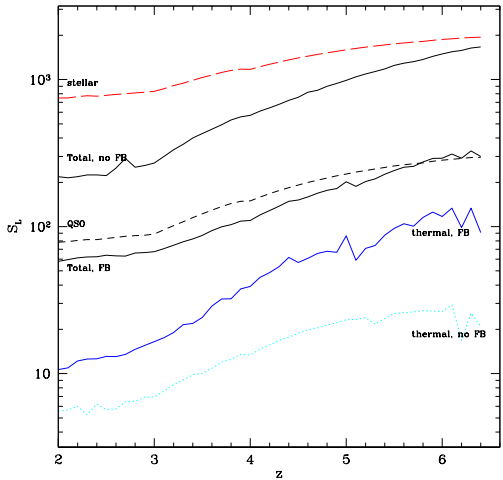}
      \caption{Redshift evolution of the softness ratio $S_L$ for 
different emission cases: thermal, no feedback (dot);
thermal, with feedback (thin solid); stellar (long dash); 
QSOs (short dash); total, no feedback (higher thick solid);
total, with feedback (lower thick solid)
              }
         \label{soft.fig}
   \end{figure}
%
%
\label{softpar.se}
Both for the individual spectral components and for the total
spectrum, we have computed the redshift evolution of the spectral
softness parameter, $S_L\equiv {J_\nHI \over J_\nHeII}$,
where $J_\nHI$ and $J_\nHeII$ are the UVB intensities at the \HI and
\HeII Lyman limits respectively. These are shown in Fig.
\ref{soft.fig}. Thermal emission are characterized
by moderate values of $S_L$, raising with redshift from $\sim 10$ to 100
and 80 to 300, respectively; the stellar component
shows less evolution but with maximal $S_L$ values above 1000 at $z \simgt
3$. The composite spectra (feedback case) $S_L$ evolution resembles
very closely that of QSOs, even at high redshift: this is somewhat
fortuitous as at $z\simgt 4$ the UVB is dominated by the sum of stellar
and thermal contribution.  When compared with the
available data, \eg the recent study of \citet{Heap00},
our predicted values for $S_L$ at redshift around $z=3.2$
are close to the measurements of those authors. However, 
the reported change in the parameter $S$ would not necessarily 
imply a ``jump'' of the same quantity at $z\simeq 3$, as it would be 
due to the evolution of our composite spectrum; 
in particular to the decrease in the ratio of thermal to the QSO flux. 
%
%
%
%
%
\subsection{\HeII Reionization} \label{heiireion.se}
The previous results hint at the intriguing possibility that \HeII
reionization could have been powered by UV light from cosmic structure
formation. For this to be the case,
the production rate of ionizing photons has to satisfy the
condition
\begin{equation}
\Gamma \times \min(\tau_{rec},\tau_{Hubble}) \geq 1 .
\label{reccon}
\end{equation}
Here $\tau_{rec} \simeq 0.9/ \bar n_{gas} \alpha(T) C $ is the
recombination time, $\alpha$ is the radiative recombination coefficient,
and the clumping factor $C \equiv \langle n_p^2\rangle/\langle
n_p\rangle ^2 > 1$ is meant to allow for the effects of density
inhomogeneities inside the ionized region.  Using a helium
to hydrogen number ratio $y=0.08$ and assuming a temperature of the
reionized gas $T \approx 4 \times 10^4$~K,  we find
\begin{equation}
\tau_{rec} = 5.3 \times 10^{15}\; C^{-1} \left(1+z\over 10\right)^{-3} \; {\rm ~s},
\label{trecn}
\end{equation}
which is shorter than a Hubble time for $z \gsim 4.5$.  Thus, from
Fig. \ref{gammi.fig} we find that at $z\approx 6$ thermal emission
dominates the photoionization rates and alone provides $5.5\times
10^{-17}$ ($10^{-16}$) \HeII photoionizations/s in the no-feedback
(feedback) case. According to eq. (\ref{trecn}), the
\HeII recombination rate at the same redshift is $6.4\times 10^{-17}
C$~s$^{-1}$. Hence,
it appears that structure formation can produce \HeII reionization
around $z=6$, without the contribution from any other process and
essentially independently of the feedback prescription adopted.

Whether or not this possibility is fully compatible with all 
observational results is not clear at the moment and will be 
the subject of future investigations. 

\section{Summary}

We have shown that UVB ionizing photons
can be copiously produced by thermal emission from shock-heated gas
in collapsing cosmic structures. 

Thermal radiation is characterized by
a hard spectrum extending up to photon energies of order $h_p \nu\sim
k_B T$. This is well above the \HI and \HeII ionization thresholds for
virial temperatures above $10^5$ K.
The bulk of the emission is produced by halos
with temperatures between 10$^6$ K and a few $\times 10^7$ K,
corresponding to masses $10^{11-13}$ M$_\odot$. 
We assume that most of the thermal radiation is able to freely
escape into intergalactic space, which is justified for a gas
that is collisionally ionized and at the temperature of these halos
\citep{mfwb04}.

We use simplified radiative transfer to compute the transmitted flux
due to QSO, stellar and thermal emissions.  Importantly, the resulting
associated photoionization rates, when compared to measurements of the
Lyman series Gunn-Peterson effect in the spectra of high redshift QSOs
\citep{FanAJ2002,BeckerAJ2001}, imply an escape fraction of UV
ionizing photons from galaxies, $f_{esc}$, below a few \%. This result
is in agreement with very recent and independent determinations of
$f_{esc}$ carried out by \citet{soto03}, who set a 3$\sigma$
(statistical) upper limit $f_{esc} \simlt 4$\% for a sample of
spectroscopically identified galaxies of redshift $1.9 < z < 3.5$ in
the Hubble Deep Field.

With $f_{esc}\simeq 1\%$, it turns out that near the
\HI ionization threshold, thermal emission is comparable to the
stellar component and amounts to about 5-10 \%, 15-30 \% and
20-50 \% of the total at redshifts of 3, 4.5 and higher
respectively. 
Near the ionization threshold for
\HeII, the thermal contribution is much stronger. It is comparable to
the QSO input already at $z\sim 3$, and it dominates for
$z> 4$. Thus, this contribution, with a
typical softness parameter $S_L=10-100$, is expected
to play a major role in \HeII reionization. 
In principle
structure formation alone provides enough photons to produce
and sustain \HeII reionization at $z\sim 6$. 

\begin{acknowledgements}
I am grateful to the organizers for the hospitality and financial support
and to G. De Lucia for reading the manuscript.
This work was partially supported by the Research
and Training Network `The Physics of the Intergalactic Medium',
EU contract HPRN-CT2000-00126 RG29185.
\end{acknowledgements}

\bibliographystyle{aa}
\bibliography{papers,books,papigm}

\begin{thebibliography}{36}
\expandafter\ifx\csname natexlab\endcsname\relax\def\natexlab#1{#1}\fi

\bibitem[{{Abel} \& {Haehnelt}(1999)}]{AbelApJ1999}
{Abel}, T. \& {Haehnelt}, M.~G. 1999, \apjl, 520, L13

\bibitem[{{Becker} {et~al.}(2001){Becker}, {Fan}, {White}, {Strauss},
  {Narayanan}, {Lupton}, {Gunn}, {Annis}, {Bahcall}, {Brinkmann}, {Connolly},
  {Csabai}, {Czarapata}, {Doi}, {Heckman}, {Hennessy}, {Ivezi{\' c}}, {Knapp},
  {Lamb}, {McKay}, {Munn}, {Nash}, {Nichol}, {Pier}, {Richards}, {Schneider},
  {Stoughton}, {Szalay}, {Thakar}, \& {York}}]{BeckerAJ2001}
{Becker}, R.~H., {Fan}, X., {White}, R.~L., {et~al.} 2001, AJ, 122, 2850

\bibitem[{{Bianchi} {et~al.}(2001){Bianchi}, {Cristiani}, \&
  {Kim}}]{BianchiA&A2001}
{Bianchi}, S., {Cristiani}, S., \& {Kim}, T.-S. 2001, A\&A, 376, 1

\bibitem[{Boksenberg {et~al.}(2003)Boksenberg, Sargent, \& Rauch}]{bosara03}
Boksenberg, A., Sargent, W.~L., \& Rauch, M. 2003, \apjs, submitted, eprint
  arXiv:astro-ph/0307557

\bibitem[{{Boyle} {et~al.}(2000){Boyle}, {Shanks}, {Croom}, {Smith}, {Miller},
  {Loaring}, \& {Heymans}}]{BoyleMNRAS2000}
{Boyle}, B.~J., {Shanks}, T., {Croom}, S.~M., {et~al.} 2000, MNRAS, 317, 1014

\bibitem[{{Boyle} {et~al.}(1988){Boyle}, {Shanks}, \&
  {Peterson}}]{BoyleMNRAS1988}
{Boyle}, B.~J., {Shanks}, T., \& {Peterson}, B.~A. 1988, MNRAS, 235, 935

\bibitem[{{Bruzual A.} \& {Charlot}(1993)}]{BruzualApJ1993}
{Bruzual A.}, G. \& {Charlot}, S. 1993, \apj, 405, 538

\bibitem[{{Bryan} {et~al.}(1999){Bryan}, {Machacek}, {Anninos}, \&
  {Norman}}]{BryanApJ1999}
{Bryan}, G.~L., {Machacek}, M., {Anninos}, P., \& {Norman}, M.~L. 1999, \apj,
  517, 13

\bibitem[{{Cen} {et~al.}(1994){Cen}, {Miralda-Escude}, {Ostriker}, \&
  {Rauch}}]{CenApJL1994}
{Cen}, R., {Miralda-Escude}, J., {Ostriker}, J.~P., \& {Rauch}, M. 1994, \apjl,
  437, L9

\bibitem[{{Fan} {et~al.}(2002){Fan}, {Narayanan}, {Strauss}, {White}, {Becker},
  {Pentericci}, \& {Rix}}]{FanAJ2002}
{Fan}, X., {Narayanan}, V.~K., {Strauss}, M.~A., {et~al.} 2002, AJ, 123, 1247

\bibitem[{{Fan} {et~al.}(2001){Fan}, {Strauss}, {Schneider}, {Gunn}, {Lupton},
  {Becker}, {Davis}, {Newman}, {Richards}, {White}, {Anderson}, {Annis},
  {Bahcall}, {Brunner}, {Csabai}, {Hennessy}, {Hindsley}, {Fukugita}, {Kunszt},
  {Ivezic}, {Knapp}, {McKay}, {Munn}, {Pier}, {Szalay}, \& {York}}]{FanAJ2001}
{Fan}, X., {Strauss}, M., {Schneider}, D., {et~al.} 2001, AJ, 121, 31

\bibitem[{Fern\'andez-Soto {et~al.}(2003)Fern\'andez-Soto, Lanzetta, \&
  Chen}]{soto03}
Fern\'andez-Soto, A., Lanzetta, K.~M., \& Chen, H.-W. 2003, \mnras, submitted,
  preprint: astro-ph/0303286

\bibitem[{Haardt \& Madau(1996)}]{hama96}
Haardt, F. \& Madau, P. 1996, \apj, 461, 20

\bibitem[{Heap {et~al.}(2000)Heap, Williger, Smette, Hubeny, Meena, Jenkins, \&
  nd~J.~N.~Winkler}]{Heap00}
Heap, S.~R., Williger, G.~M., Smette, A., {et~al.} 2000, \apj, 534, 69

\bibitem[{Kauffmann {et~al.}(2003)Kauffmann, Heckman, White, Charlot, Tremonti,
  Brinchmann, \& et~al.}]{kauffmannetal03}
Kauffmann, G., Heckman, T.~M., White, S. D.~M., {et~al.} 2003, \mnras, 341, 33

\bibitem[{{Kim} {et~al.}(2001){Kim}, {Cristiani}, \& {D'Odorico}}]{KimA&A2001}
{Kim}, T.~S., {Cristiani}, S., \& {D'Odorico}, S. 2001, A\&A, 373, 757

\bibitem[{{Kim} {et~al.}(2002){Kim}, {Cristiani}, \& {D'Odorico}}]{KimA&A2002}
{Kim}, T.~S., {Cristiani}, S., \& {D'Odorico}, S. 2002, \aap, 383, 747

\bibitem[{{Madau} \& {Efstathiou}(1999)}]{MadauApJL1999}
{Madau}, P. \& {Efstathiou}, G. 1999, \apjl, 517, L9

\bibitem[{{Madau} {et~al.}(1998){Madau}, {Pozzetti}, \&
  {Dickinson}}]{MadauApJ1998}
{Madau}, P., {Pozzetti}, L., \& {Dickinson}, M. 1998, ApJ, 498, 106

\bibitem[{{McDonald} \& {Miralda-Escud{\' e}}(2001)}]{McDonald2001ApJ}
{McDonald}, P. \& {Miralda-Escud{\' e}}, J. 2001, \apjl, 549, L11

\bibitem[{Miniati {et~al.}(2004)Miniati, Ferrara, White, \& Bianchi}]{mfwb04}
Miniati, F., Ferrara, A., White, S. D.~M., \& Bianchi, S. 2004, \mnras, in
  press, e-print astro-ph/0309127

\bibitem[{{Miralda-Escude} {et~al.}(1996){Miralda-Escude}, {Cen}, {Ostriker},
  \& {Rauch}}]{MiraldaEscudeApJ1996}
{Miralda-Escude}, J., {Cen}, R., {Ostriker}, J.~P., \& {Rauch}, M. 1996, \apj,
  471, 582

\bibitem[{{Nath} {et~al.}(1999){Nath}, {Sethi}, \&
  {Shchekinov}}]{NathMNRAS1999}
{Nath}, B.~B., {Sethi}, S.~K., \& {Shchekinov}, Y. 1999, MNRAS, 303, 1

\bibitem[{{Paresce} {et~al.}(1980){Paresce}, {McKee}, \&
  {Bowyer}}]{ParesceApJ1980}
{Paresce}, F., {McKee}, C.~F., \& {Bowyer}, S. 1980, ApJ, 240, 387

\bibitem[{Peebles(1993)}]{peebles93}
Peebles, P. J.~E. 1993, Principles of Physical Cosmology (Princeton New Jersey:
  Princeton University Press)

\bibitem[{Raymond \& Smith(1977)}]{rasm77}
Raymond, J.~C. \& Smith, B.~W. 1977, \apjs, 35, 419

\bibitem[{{Schirber} \& {Bullock}(2002)}]{SchirberApJ2002}
{Schirber}, M. \& {Bullock}, J.~S. 2002, ApJ, 584, 110

\bibitem[{{Scott} {et~al.}(2000){Scott}, {Bechtold}, {Dobrzycki}, \& {Kulkar
  ni}}]{ScottApJS2000}
{Scott}, J., {Bechtold}, J., {Dobrzycki}, A., \& {Kulkar ni}, V.~P. 2000, ApJS,
  130, 67

\bibitem[{Sheth \& Tormen(1999)}]{shto99}
Sheth, R.~K. \& Tormen, G. 1999, \mnras, 308, 119

\bibitem[{Songaila(1998)}]{Songaila1998}
Songaila, A. 1998, \aj, 115, 2184

\bibitem[{{Steidel} {et~al.}(1999){Steidel}, {Adelberger}, {Giavalisco},
  {Dickinson}, \& {Pettini}}]{SteidelApJ1999}
{Steidel}, C.~C., {Adelberger}, K.~L., {Giavalisco}, M., {Dickinson}, M., \&
  {Pettini}, M. 1999, ApJ, 519, 1

\bibitem[{{Theuns} {et~al.}(1998){Theuns}, {Leonard}, \&
  {Efstathiou}}]{TheunsMNRAS1998a}
{Theuns}, T., {Leonard}, A., \& {Efstathiou}, G. 1998, \mnras, 297, L49

\bibitem[{White \& Rees(1978)}]{whre78}
White, S. D.~M. \& Rees, M.~J. 1978, \mnras, 183, 341

\bibitem[{{Zhang} {et~al.}(1995){Zhang}, {Anninos}, \& {Norman}}]{ZhangApJ1995}
{Zhang}, Y., {Anninos}, P., \& {Norman}, M.~L. 1995, \apjl, 453, L57

\bibitem[{{Zheng} {et~al.}(1997){Zheng}, {Kriss}, {Telfer}, {Grimes}, \&
  {Davidsen}}]{ZhengApJ1997}
{Zheng}, W., {Kriss}, G.~A., {Telfer}, R.~C., {Grimes}, J.~P., \& {Davidsen},
  A.~F. 1997, ApJ, 475, 469

\bibitem[{{Zuo}(1993)}]{ZuoA&A1993}
{Zuo}, L. 1993, A\&A, 278, 343

\end{thebibliography}

\end{document}